\documentclass[twocolumn]{article}

\usepackage[a4paper,textwidth=516pt,columnsep=12pt,top=20mm,bottom=25mm]{geometry}
\usepackage{graphicx}
\usepackage{amsmath,amsfonts}
\usepackage{amssymb}
\usepackage{natbib}   
\usepackage{booktabs}
\usepackage{bm}
\usepackage{algorithm}
\usepackage{algpseudocode}

\newtheorem{remark}{Remark}
\newtheorem{prob}{Problem}
\newtheorem{ass}{Assumption}

\newcommand{\Real}{\mathbf{R}}

\newcommand{\diag}{\mathop{\mathrm{diag}}}

\newcommand{\col}{\mathop{\mathrm{col}}}

\newcommand{\argmin}{\mathop{\mathrm{arg\,min}}}

\begin{document}

\title{Iterative Identification of Aperiodic Sampled-Data Systems via Affine LPV Models Scheduled by the Sampling Interval\thanks{Corresponding author H.~Okajima.}}

\author{%
  Hiroshi Okajima\\
  {\normalsize Faculty of Advanced Science and Technology}\\
  {\normalsize Kumamoto University, Kumamoto, Japan}\\
  {\normalsize\texttt{okajima@mech.kumamoto-u.ac.jp}}
  \and
  Yohei Hosoe\\
  {\normalsize Graduate School of Engineering}\\
  {\normalsize Kyoto University, Kyoto, Japan}\\
  {\normalsize\texttt{hosoe.yohei.5v@kyoto-u.ac.jp}}%
}

\date{}

\maketitle

\begin{abstract}
This paper proposes a method for identifying a control-oriented model of a
stable continuous-time linear system from aperiodically sampled
input-output data.  Rather than recovering the continuous-time matrices,
the method directly identifies a discrete-time linear parameter-varying
(LPV) state-space model that depends affinely on the measured normalized
sampling-interval variation, a structure obtained from the first-order
expansion of the zero-order-hold discretization.  The resulting model is
directly usable in control design for aperiodic sampled-data systems.  An
initial LPV estimate, including approximate first-order terms, is derived
from a standard LTI subspace model through sampled-data relations.  The
method then iteratively alternates state-sequence reconstruction and LPV
parameter refitting.
Numerical examples show that the identified model retains high validation
accuracy over a wide range of sampling-interval variations where a fixed
LTI model degrades.  Already at initialization, the state-space LPV
structure substantially outperforms an LPV-ARX regression with the
same scheduling variable, and the iterative refinement brings the model
close to a state-measured benchmark, a practical upper-bound reference.
\end{abstract}

\section{Introduction}\label{sec:intro}

In many networked, embedded, and sensing-limited control systems, data are
not obtained at a single uniform sampling period.  Sampling instants may
fluctuate because of scheduling jitter, communication constraints,
event-driven measurements, or deliberately non-uniform measurement policies.
When a continuous-time plant is driven by a zero-order-hold input and
sampled at varying intervals $h_k$, the exact discrete-time
state-transition and input matrices are functions of $h_k$, so that a
fixed discrete-time model cannot accurately capture the resulting
sampled-data dynamics.  Stability and control of such aperiodic
sampled-data systems have been studied extensively; see \citet{hetel2017}
for a comprehensive review and, e.g., \citet{fujioka2009,cervin2003} for
representative results.  On the control-design side, several approaches
incorporate the sampling-interval dependence of the discrete-time matrices
directly into the model structure.  A sampling-period-dependent polytopic
LPV model has been used for $H_\infty$ controller design
\citep{robert2010}, and robust LMI conditions that treat the
sampling-interval-dependent state-transition matrix as a parametric
uncertainty have been developed in \citet{oishi2010}.  When the sampling
interval is instead generated by a stochastic process, a model of this
kind that is affine in the interval variation becomes a discrete-time
system with a random multiplicative parameter.  Its second-moment
stability can be analyzed by LMI-based conditions \citep{hosoe2019,hosoe2022}, and
it belongs to the model class assumed in stochastic MPC with
multiplicative uncertainty \citep{cannon2009}.  Moreover, a second-moment
stabilizing gain may serve as the pre-stabilizing feedback
\citep{moussa2026} in the closed-loop prediction paradigm on which
tube-based stochastic MPC \citep{cannon2011} is built.

Identification of such systems from aperiodically sampled data has also
been studied.  Existing methods fall into two groups.  The first recovers
a continuous-time model \citep{larsson2007,mu2015,chen2017,huang2023},
from which the sampling-interval-dependent discrete-time model must then
be derived as in the control-design approaches above.  This indirect route
requires a model conversion before the model can be used for stability
analysis or controller synthesis.  The second consists of lifting-based
methods for non-uniform but periodic sampling patterns
\citep{li2006,ding2009,xie2010}, which presuppose a repeating pattern of
sampling instants.  To the best of the authors' knowledge, the direct
identification of a sampling-interval-dependent discrete-time model from
aperiodically sampled data has not been addressed, although it would make
the model conversion unnecessary.

The present paper approaches this problem from the standpoint of
control-oriented identification of aperiodic sampled-data systems.  Rather
than recovering the continuous-time matrices, we directly identify a
discrete-time affine LPV state-space model in which the normalized
sampling-interval variation serves as a known scheduling variable, that
is, a model of the same class as those used in the control syntheses above.
In this way, the affine dependence on the scheduling variable is
inherited from the first-order expansion of the zero-order-hold
discretization rather than postulated, so that the LPV structure carries
a physical sampled-data interpretation.  Moreover, the identified model
can be used directly in LMI-based stability analysis and controller
synthesis \citep{robert2010,oishi2010} and, under stochastic sampling,
in second-moment stabilization \citep{hosoe2019,hosoe2022} and stochastic MPC with
multiplicative uncertainty \citep{cannon2009,moussa2026}.

The contributions of this paper are as follows.  First, the
identification problem is formulated directly in terms of the affine LPV
state-space model motivated above.  Second, an initial LPV model is
derived from a standard LTI subspace model through the sampled-data
relations between the discrete-time matrices and their first-order
sampling-interval sensitivities, with the matrix logarithm replaced by
its bilinear approximation.  Third, an iterative algorithm alternating
regularized least-squares state reconstruction and least-squares
refitting of the LPV matrices is proposed.  Finally, numerical examples
quantify the effect of the sampling-interval variation amplitude and the
sensitivity to measurement noise in comparison with a fixed LTI model, an
LPV-ARX baseline that is affine in the same scheduling variable, and a
continuous-time indirect method.

The remainder of the paper is organized as follows.
Section~\ref{sec:formulation} derives the affine LPV structure from the
first-order expansion of the zero-order-hold discretization, states the
conditions under which this approximation is reasonable, and formulates
the identification problem.  Section~\ref{sec:continuous} derives the
initial LPV model from an LTI subspace model and proposes the iterative
algorithm alternating state smoothing and LPV refitting.
Section~\ref{sec:numerical} gives numerical examples, and
Section~\ref{sec:conclusion} concludes the paper.

Notation.  $\|\cdot\|_2$ denotes the Euclidean norm of a vector
and the induced $2$-norm (largest singular value) of a matrix,
$\|\cdot\|_F$ the Frobenius norm, $\diag(\cdot)$ the diagonal matrix with
the listed entries, and $\col(\cdot)$ the column vector obtained by
stacking its arguments.

\section{Problem Formulation}\label{sec:formulation}

\subsection{Continuous-time plant and aperiodic sampling}

Consider the continuous-time linear time-invariant plant
\begin{align}
    \dot{x}(t) &= A_{\rm c}x(t)+B_{\rm c}\bigl(u(t)+\eta(t)\bigr), \label{eq:ct_state}\\
    y(t) &= C_{\rm c}x(t)+D_{\rm c}\bigl(u(t)+\eta(t)\bigr)+v(t), \label{eq:ct_output}
\end{align}
where $x(t)\in\Real^{n_x}$ is the state, $y(t)\in\Real^{n_y}$ is the
measured output, $u(t)\in\Real^{n_u}$ is the known input,
$\eta(t)\in\Real^{n_u}$ is an unmeasured input disturbance, such as an
actuator disturbance or an unmodeled plant input, and
$v(t)\in\Real^{n_y}$ is measurement noise.  No stochastic model is
imposed on $\eta(t)$ and $v(t)$.  Throughout this paper, the plant
\eqref{eq:ct_state}--\eqref{eq:ct_output} is assumed to be stable, that
is, $A_{\rm c}$ is Hurwitz, since the proposed method is based on
open-loop identification.
The output is sampled at the aperiodic sampling instants
\begin{align}
    0=t_0<t_1<t_2<\cdots<t_N,
\end{align}
with the sampling intervals
\begin{align}
    h_k:=t_{k+1}-t_k,\qquad k=0,1,\ldots,N-1. \label{eq:hk_def}
\end{align}
The known input is generated by a zero-order hold, and the disturbance
is assumed to be held over the same intervals:
\begin{align}
    u(t)=u_k,\qquad \eta(t)=\eta_k,\qquad t\in[t_k,t_{k+1}),
\end{align}
with $u_k,\eta_k\in\Real^{n_u}$.  The output sample recorded at $t_k$
is $y_k:=y(t_k)$; we write $v_k:=v(t_k)$ and $x_k:=x(t_k)$.
Integrating \eqref{eq:ct_state} over $[t_k,t_{k+1})$ gives the exact
variable-sampling discrete-time representation of the plant,
\begin{align}
    x_{k+1} &= A_{\rm d}(h_k)x_k+B_{\rm d}(h_k)(u_k+\eta_k), \label{eq:exact_varying_dt}\\
    y_k &= Cx_k+D(u_k+\eta_k)+v_k,\quad C:=C_{\rm c},\ D:=D_{\rm c}, \label{eq:exact_varying_out}
\end{align}
for $k=0,1,\ldots,N-1$, where
\begin{align}
    A_{\rm d}(h_k) &= \exp(A_{\rm c}h_k),\label{eq:Ad_exact}\\
    B_{\rm d}(h_k) &= \int_0^{h_k}\exp(A_{\rm c}\tau)B_{\rm c}\,d\tau.\label{eq:Bd_exact}
\end{align}
Since $\eta(t)$ is held over the same intervals as $u(t)$, the input and
the disturbance enter through the same matrix $B_{\rm d}(h_k)$.

For identification, the available data are the recorded sequences
$(u_k,y_k,h_k)_{k=0}^{N-1}$, where $h_k$ is computed from the recorded
instants $t_k$ and $t_{k+1}$ by \eqref{eq:hk_def}; the state $x_k$ and
the disturbances $\eta_k$ and $v_k$ are not available.  Here $N$ denotes
the number of sampling intervals, the sampling instants are indexed by
$k=0,1,\ldots,N$, and the state sequence $(x_k)_{k=0}^{N}$ contains
$N+1$ samples while $u_k$, $y_k$, and $h_k$ are indexed over
$k=0,\ldots,N-1$.  The terminal output sample $y_N$, if available, is not
used in the identification.  Since the identification is carried out in
batch from recorded data, $h_k=t_{k+1}-t_k$ need not be known before time
$t_{k+1}$.

\subsection{Sampling-interval-dependent LPV approximation}

Let $\bar{h}>0$ be a nominal sampling interval, that is, a fixed value
within the observed range $[h_{\min},h_{\max}]$ of the varying intervals
$h_k$, around which their variation is measured.  We
define the normalized sampling-interval variation by
\begin{align}
    \delta_k := \frac{h_k-\bar{h}}{\bar{h}},\qquad k=0,1,\ldots,N-1. \label{eq:delta_def}
\end{align}
Any choice of $\bar{h}$ that keeps $|\delta_k|$ small for all $k$ is
admissible.  Typical choices are the sample mean, the median, and the
midpoint $(h_{\max}+h_{\min})/2$, which conveniently center $\delta_k$
around zero.  For moderate variations of $h_k$, the
exact matrices
$A_{\rm d}(h_k)$ and $B_{\rm d}(h_k)$ in
\eqref{eq:Ad_exact}--\eqref{eq:Bd_exact} are approximated by affine
functions of $\delta_k$:
\begin{align}
    A_{\rm d}(h_k) &\simeq A_0+A_1\delta_k,\label{eq:A_affine}\\
    B_{\rm d}(h_k) &\simeq B_0+B_1\delta_k.\label{eq:B_affine}
\end{align}
This approximation is motivated by the Taylor expansion around
$h=\bar{h}$, a standard technique in the control-design literature
\citep[see, e.g.,][]{robert2010,oishi2010}.  Since $h_k=\bar{h}(1+\delta_k)$ and the matrices $A_{\rm c}\bar{h}$ and
$A_{\rm c}\bar{h}\delta_k$ commute (both being scalar multiples of $A_{\rm c}$),
the matrix exponential factorizes as
$\exp(A_{\rm c}\bar{h}(1+\delta_k))=\exp(A_{\rm c}\bar{h})\exp(A_{\rm c}\bar{h}\delta_k)$
and admits the expansion
\begin{align}
    A_{\rm d}(h_k)
    &= \exp(A_{\rm c}\bar{h}(1+\delta_k)) \notag\\
    &= \exp(A_{\rm c}\bar{h})
    +\bar{h}A_{\rm c}\exp(A_{\rm c}\bar{h})\delta_k+O(\delta_k^2).
    \label{eq:Ad_taylor}
\end{align}
Analogously, $dB_{\rm d}(h)/dh=\exp(A_{\rm c}h)B_{\rm c}$ together with
$dh_k/d\delta_k=\bar{h}$ gives
\begin{align}
    B_{\rm d}(h_k)
    = B_{\rm d}(\bar{h})+\bar{h}\exp(A_{\rm c}\bar{h})B_{\rm c}\delta_k+O(\delta_k^2).
    \label{eq:Bd_taylor}
\end{align}
The affine coefficients are thus
$A_0=\exp(A_{\rm c}\bar{h})$, $A_1=\bar{h}A_{\rm c}\exp(A_{\rm c}\bar{h})$,
$B_0=B_{\rm d}(\bar{h})$, and $B_1=\bar{h}\exp(A_{\rm c}\bar{h})B_{\rm c}$.  Since a
state-space realization is defined only up to a similarity transformation,
these expressions serve as structural motivation for the affine
parametrization rather than as quantities to be computed from $A_{\rm c}$.
In the identification, $A_0,A_1,B_0,B_1$ are treated as
independent discrete-time matrices, and $A_{\rm c}$ need not be recovered.
The approximation accuracy depends on $\|A_{\rm c}\bar{h}\|_2$ and the range of
$\delta_k$, and is assessed through validation performance rather than
realization-dependent matrix-norm errors.

The scheduling variable
$\delta_k$ is treated as a known exogenous signal computed from the measured
sampling interval $h_k$; no stochastic model is imposed on the sequence
$(\delta_k)_{k=0}^{N-1}$.

We summarize the conditions under which the affine approximation is
reasonable as the following assumption.

\begin{ass}\label{ass:plant_sampling}
Let $\delta_{\max}:=\max_{0\le k\le N-1}|\delta_k|$ denote the normalized
sampling-interval variation amplitude over the data record.
The product $\|A_{\rm c}\|_2 \bar{h}\delta_{\max}$ is small enough that the
second-order Taylor remainders in \eqref{eq:Ad_taylor}--\eqref{eq:Bd_taylor}
are negligible.
\end{ass}
Assumption~\ref{ass:plant_sampling} concerns the range of validity of the affine model.
Since $\bar{h}\delta_{\max}=\max_{0\le k\le N-1}|h_k-\bar{h}|$ and
the second-order term is of order
$O(\|A_{\rm c}\bar{h}\|_2^2\delta_{\max}^2)=O(\|A_{\rm c}\|_2^2(\bar{h}\delta_{\max})^2)$,
the approximation accuracy depends on $\bar{h}$ only through the maximum
distance from $\bar{h}$ to the observed sampling intervals, which is minimized
by the midpoint of $[h_{\min},h_{\max}]$.

Substituting the affine approximation
\eqref{eq:A_affine}--\eqref{eq:B_affine} into
\eqref{eq:exact_varying_dt}--\eqref{eq:exact_varying_out} gives the
relations satisfied by the recorded sequences
$(u_k,y_k,h_k)_{k=0}^{N-1}$ together with the unmeasured $x_k$,
$\eta_k$, and $v_k$:
\begin{align}
    x_{k+1} &= (A_0+A_1\delta_k)x_k+(B_0+B_1\delta_k)(u_k+\eta_k)
    +e_k,
    \label{eq:lpv_state}\\
    y_k &= Cx_k+D(u_k+\eta_k)+v_k,\qquad k=0,1,\ldots,N-1,\label{eq:lpv_output}
\end{align}
where $e_k$ is the modeling error due to the affine
approximation, that is, the structured truncation error
$[A_{\rm d}(h_k)-(A_0+A_1\delta_k)]x_k+[B_{\rm d}(h_k)-(B_0+B_1\delta_k)](u_k+\eta_k)$,
which depends on the state, the input, the input disturbance, and the
sampling interval.  When $\eta_k=0$, $v_k=0$, and $e_k=0$,
\eqref{eq:lpv_state}--\eqref{eq:lpv_output} reduce to an LPV model
driven only by the known signals $u_k$ and $\delta_k$.

The identification objective is to estimate this LPV model, namely
\begin{align}
    x_{k+1} &= (A_0+A_1\delta_k)x_k+(B_0+B_1\delta_k)u_k,
    \label{eq:lpv_state_det}\\
    y_k &= Cx_k+Du_k,\label{eq:lpv_output_det}
\end{align}
from the sampling intervals and the input-output data generated under the
disturbances $\eta_k$ and $v_k$.  The problem is stated as follows.

\begin{prob}\label{prob:id}
Let the data $(u_k,y_k,h_k)_{k=0}^{N-1}$ be generated by the plant
\eqref{eq:ct_state}--\eqref{eq:ct_output} under the aperiodic sampling
\eqref{eq:hk_def}, so that they satisfy the exact representation
\eqref{eq:exact_varying_dt}--\eqref{eq:exact_varying_out}, and let
Assumption~\ref{ass:plant_sampling} hold.  Given these data, the
scheduling variable $\delta_k$ computed from $h_k$ by
\eqref{eq:delta_def}, and the model order $n_x$, find the matrices
$A_0,A_1,B_0,B_1,C,D$ of the sampling-interval-dependent LPV model
\eqref{eq:lpv_state_det}--\eqref{eq:lpv_output_det} such that the
free-run output of the model driven by $u_k$ and $\delta_k$ approximates
the measured output $y_k$.  Since the model is a state-space
realization, the matrices are sought up to a similarity transformation.
\end{prob}

Problem~\ref{prob:id} is not a generic LPV identification problem with an
arbitrary scheduling signal and arbitrary basis functions.  The plant is
linear time-invariant, the scheduling variable $\delta_k$ is computed by
\eqref{eq:delta_def} from the given sampling intervals $h_k$, and the
$\delta_k$-terms are induced purely by the variable-sampling
discretization through the first-order expansion
\eqref{eq:Ad_taylor}--\eqref{eq:Bd_taylor}.  Equivalently, the one-step
contributions of the state and input are
\begin{align}
    A_{\rm d}(h_k)x_k
    &\simeq A_0x_k + A_1(\delta_k x_k),\label{eq:Ad_first_order_signal}\\
    B_{\rm d}(h_k)u_k
    &\simeq B_0u_k + B_1(\delta_k u_k),\label{eq:Bd_first_order_signal}
\end{align}
so the products $\delta_k x_k$ and $\delta_k u_k$ are the first-order
sampling-interval-dependent components of the sampled-data representation,
not regressors chosen by the designer as in a general-purpose LPV
parametrization.  Of these quantities, $\delta_k$ and
$\delta_k u_k$ are known exogenous regressors computable in advance, whereas
$\delta_k x_k$ depends on the unmeasured state and becomes available only
after a state trajectory is reconstructed.

In Problem~\ref{prob:id}, the state sequence $(x_k)_{k=0}^{N}$ is not measured.
If it were available, the above structure would lead to a linear
least-squares problem because the unknown matrices enter linearly once the
products $\delta_k x_k$ and $\delta_k u_k$ are regarded as known data; the
corresponding regression equation is
\begin{align}
    X_+ &=
    \begin{bmatrix} A_0&A_1&B_0&B_1\end{bmatrix}
    \begin{bmatrix}
        X\\
        X\Delta\\
        U\\
        U\Delta
    \end{bmatrix},\label{eq:state_measured_ls}
\end{align}
where
\begin{align}
    X&=\begin{bmatrix}x_0&x_1&\cdots&x_{N-1}\end{bmatrix},\\
    X_+&=\begin{bmatrix}x_1&x_2&\cdots&x_N\end{bmatrix},\\
    U&=\begin{bmatrix}u_0&u_1&\cdots&u_{N-1}\end{bmatrix},\\
    \Delta&=\diag(\delta_0,\delta_1,\ldots,\delta_{N-1}).
\end{align}
Here, $X\Delta$ and $U\Delta$ denote columnwise multiplication of the state
and input data by the corresponding scheduling values $\delta_k$.  The main
difficulty is therefore not the estimation of the LPV coefficient matrices
once the state trajectory is known, but the reconstruction, from
input-output data only, of a state trajectory that is compatible with the
sampling-interval-dependent state equation.

The identified model is intended for direct use in control design: the
matrices $A_0,A_1,B_0,B_1,C,D$ are used as they are, without the matrix
logarithm, re-discretization, or other model conversion required by
indirect routes that first estimate $(A_{\rm c},B_{\rm c})$.  How
the model enters control design depends on the information on $h_k$
available at time $t_k$, as summarized in Remarks~\ref{rem:det_use}
and~\ref{rem:stoch_use}.

\begin{remark}\label{rem:det_use}
If $h_k$ is available at time $t_k$, for example when the next sampling
instant is assigned by a scheduler, the identified model
\eqref{eq:lpv_state_det}--\eqref{eq:lpv_output_det} is an LPV model with a
measured scheduling variable, to which LMI-based gain-scheduled designs
such as \citet{robert2010} apply.  If $h_k$ is not available at time $t_k$
but is confined to a known interval, then, since the model is affine in the
single scheduling variable $\delta_k$, evaluating it at the endpoints of
that interval gives the two vertex systems of a polytopic model, to which
robust LMI conditions such as \citet{oishi2010} apply.
\end{remark}

\begin{remark}\label{rem:stoch_use}
If $h_k$ is generated by a stochastic process with known distribution or
known moments, $\delta_k$ is a random multiplicative parameter of the
identified model.  The second-moment stability of the closed loop under a
constant state-feedback gain can then be verified by LMI-based conditions
\citep{hosoe2019,hosoe2022}, and the model belongs to the class assumed in
stochastic MPC with multiplicative uncertainty \citep{cannon2009}.  A gain
verified in this way can also serve as the pre-stabilizing feedback
\citep{moussa2026} in the closed-loop prediction paradigm of tube-based
stochastic MPC \citep{cannon2011}.
\end{remark}

\section{Proposed Identification Method}\label{sec:continuous}

\subsection{LPV initialization from an LTI subspace model}
\label{sec:initial_model}

Standard LTI subspace methods such as MOESP and N4SID
\citep{verhaegen1992,overschee1994} rely on a lifted data equation with a
fixed extended observability matrix and a block-Toeplitz Markov-parameter
structure; scheduling-dependent LPV extensions exist
\citep{felici2007,verhaegen_lpv}.
For the sampling-interval-dependent model this fixed structure is lost: the
coefficient from $x_k$ to $y_{k+i}$,
\begin{align}
    C(A_0+A_1\delta_{k+i-1})\cdots(A_0+A_1\delta_k),
    \label{eq:lpv_multi_step}
\end{align}
depends on the local scheduling trajectory and varies column to column, and
the input coefficients likewise do not form a fixed Markov-parameter matrix.
Hence standard MOESP/N4SID is not a direct identification method for the
present LPV model, but it is still useful for initialization.

In the proposed method, a standard LTI subspace identification method, for
example N4SID, is applied to the input-output data while ignoring the
variation of $h_k$.  The irregularly spaced data are treated as a
uniformly sampled sequence with the nominal sampling interval
$\bar{h}$ of Section~\ref{sec:formulation}.  This step is expected to yield a
fixed discrete-time realization
\begin{align}
    x^{\rm LTI}_{k+1} &= A_{\rm LTI}x^{\rm LTI}_k+B_{\rm LTI}u_k,\label{eq:lti_initial_state}\\
    y_k &= C_{\rm LTI}x^{\rm LTI}_k+D_{\rm LTI}u_k,\label{eq:lti_initial_output}
\end{align}
capturing the dominant dynamics averaged over the observed intervals; it is
not intended to represent the sampling-interval dependence.
The zeroth-order LPV matrices are set to the LTI model directly:
\begin{align}
    A_0^{(0)} &= A_{\rm LTI}, & B_0^{(0)} &= B_{\rm LTI},\label{eq:init_AB0}\\
    C^{(0)} &= C_{\rm LTI}, & D^{(0)} &= D_{\rm LTI}.\label{eq:init_CD}
\end{align}
For the first-order matrices $A_1^{(0)}$ and $B_1^{(0)}$, a non-trivial
initial estimate can be obtained from the sampled-data relations of
Section~\ref{sec:formulation}, which link the first-order matrices to
the zeroth-order ones.  For the true plant,
\eqref{eq:Ad_taylor}--\eqref{eq:Bd_taylor} give $A_1=(A_{\rm c}\bar{h})A_0$
and, by \eqref{eq:Bd_exact},
$B_0=B_{\rm d}(\bar{h})=A_{\rm c}^{-1}(A_0-I)B_{\rm c}$, where $A_{\rm c}$
and $A_0-I$ are nonsingular because $A_{\rm c}$ is Hurwitz; hence
$B_1=\bar{h}A_0B_{\rm c}=A_0(A_{\rm c}\bar{h})(A_0-I)^{-1}B_0$, where all
factors commute as functions of $A_{\rm c}$.  Since
$A_0=\exp(A_{\rm c}\bar{h})$, these relations involve only $A_0$, $B_0$,
and a matrix logarithm $A_{\rm c}\bar{h}$ of $A_0$, and they are
preserved under similarity transformations; they can therefore be
applied to the identified pair $(A_{\rm LTI},B_{\rm LTI})$ in place of
$(A_0,B_0)$.  Rather than computing the matrix logarithm of
$A_{\rm LTI}$, we use the bilinear (Tustin) approximation
$A_{\rm c}\bar{h}\approx 2(A_0-I)(A_0+I)^{-1}$, the matrix version of the
$[1,1]$ Pad\'{e} approximant $\log z\approx 2(z-1)/(z+1)$ at $z=1$,
whose error is of third order in $A_{\rm c}\bar{h}$.  When this
approximation is substituted into the relations above, the factor
$A_0-I$ cancels $(A_0-I)^{-1}$ by commutativity; replacing $A_0$, $B_0$
by $A_{\rm LTI}$, $B_{\rm LTI}$ then gives
\begin{align}
    A_1^{(0)} &= 2(A_{\rm LTI}-I)(A_{\rm LTI}+I)^{-1}A_{\rm LTI},
    \label{eq:init_A1}\\
    B_1^{(0)} &= 2A_{\rm LTI}(A_{\rm LTI}+I)^{-1}B_{\rm LTI}.
    \label{eq:init_B1}
\end{align}
Only $(A_{\rm LTI}+I)^{-1}$ is needed, which exists whenever $A_{\rm LTI}$
is Schur stable, as expected for a stable plant; the Hurwitz property of
$A_{\rm c}$ is used only in the exact relations above.  In this way, the
initial model already carries an approximate first-order
sampling-interval dependence, derived from the LTI realization through
the sampled-data matrix relations.

\subsection{Iterative state smoothing and LPV refitting}\label{sec:iterative}

For notational convenience, let $\theta$ denote the collection of all LPV
model matrices to be identified, that is,
$\theta=(A_0,A_1,B_0,B_1,C,D)$.  The estimate of $\theta$ at iteration
$r$ is denoted by
$\theta^{(r)}=(A_0^{(r)},A_1^{(r)},B_0^{(r)},B_1^{(r)},C^{(r)},D^{(r)})$,
and its initial value $\theta^{(0)}$ is given by
\eqref{eq:init_AB0}--\eqref{eq:init_B1} of Section~\ref{sec:initial_model}.
Starting from $\theta^{(0)}$, the refinement alternates two conditional
least-squares problems: with $\theta^{(r)}$ fixed, the state trajectory is
reconstructed by smoothing, and with the new trajectory fixed, the LPV
matrices are refitted by least squares to give $\theta^{(r+1)}$.  The
procedure thus repeats
\begin{align}
    \theta^{(r)}
    \ \longrightarrow
    \hat{X}^{(r+1)}
    \ \longrightarrow
    \theta^{(r+1)}.
    \label{eq:iteration_flow}
\end{align}

\subsubsection*{State-smoothing step}

Given a current estimate of the LPV matrices at iteration $r$, denote
\begin{align}
    \mathcal{A}_k^{(r)}&:=A_0^{(r)}+A_1^{(r)}\delta_k, \notag\\
    \mathcal{B}_k^{(r)}&:=B_0^{(r)}+B_1^{(r)}\delta_k,
    \quad k=0,\ldots,N-1.
\end{align}
In this step, the superscript $(r)$ is omitted from $\mathcal{A}_k$, $\mathcal{B}_k$, $C$, and
$D$, which are all evaluated at the current iterate $\theta^{(r)}$.
The state sequence is reconstructed as the minimizer of the following
deterministic regularized least-squares criterion over the entire state
trajectory $(x_k)_{k=0}^{N}$:
\begin{align}
    \bigl(\hat{x}_k^{(r+1)}\bigr)_{k=0}^{N}
    &= \argmin_{x_0,x_1,\ldots,x_N}\Bigl\{
    \sum_{k=0}^{N-1}\|y_k-Cx_k-Du_k\|_2^2 \notag\\
    &\qquad+\alpha\sum_{k=0}^{N-1}\|x_{k+1}-\mathcal{A}_kx_k-\mathcal{B}_ku_k\|_2^2\Bigr\},
    \label{eq:state_smoothing}
\end{align}
where the first and second terms penalize the residuals of the output equation
\eqref{eq:lpv_output_det} and the state equation \eqref{eq:lpv_state_det}
evaluated at the current iterate, both of which would vanish at the true
state trajectory in the absence of $v_k$, $\eta_k$, $e_k$, and estimation
error in $\theta^{(r)}$.  The weighting parameter $\alpha>0$ balances the
two residuals: a large $\alpha$ forces the state trajectory to be
consistent with the current LPV model, while a small $\alpha$ emphasizes
fitting the measured output.  In a Gaussian state-space interpretation with isotropic errors,
$\alpha$ would correspond to a relative inverse-variance weight.  In this
paper, however, \eqref{eq:state_smoothing} is used as a deterministic
regularized criterion, because the modeling error due to affine truncation and the
input disturbance need not be independent isotropic process noise.
Therefore, $\alpha$ is selected empirically.
Since the criterion in \eqref{eq:state_smoothing} is convex quadratic in
$(x_k)_{k=0}^{N}$, the minimization is assembled in the implementation
as a linear least-squares problem
$\min_{\xi}\|b-\Psi\xi\|_2^2$ in the stacked state vector
$\xi:=\col(x_0,x_1,\ldots,x_N)$, where $\Psi$ and $b$ are built from $C$,
$D$, $\mathcal{A}_k$, $\mathcal{B}_k$, $\alpha$, and the data.  Since each residual in
\eqref{eq:state_smoothing} involves only $x_k$ and $x_{k+1}$, $\Psi$ is
sparse, and the problem is solved by a sparse direct least-squares solver.

The minimizing sequence $\bigl(\hat{x}_k^{(r+1)}\bigr)_{k=0}^{N}$ in
\eqref{eq:state_smoothing} is arranged in matrix form as
\begin{align}
    \hat{X}^{(r+1)}:=
    \begin{bmatrix}\hat{x}_0^{(r+1)} & \hat{x}_1^{(r+1)} & \cdots & \hat{x}_N^{(r+1)}\end{bmatrix},
    \label{eq:Xhat_def}
\end{align}
which serves as the reconstructed state trajectory for the parameter-refitting
step below.

\subsubsection*{Parameter-refitting step}

Given the reconstructed state trajectory $\hat{X}^{(r+1)}$ defined in
\eqref{eq:Xhat_def}, the matrices $A_0,A_1,B_0,B_1,C,D$ are re-estimated
by linear least squares.  In this step, the superscript $(r+1)$ is omitted,
so that $\hat{X}$ and $\hat{x}_k$ stand for $\hat{X}^{(r+1)}$ and
$\hat{x}_k^{(r+1)}$.  With $U$ and $\Delta$ as defined after
\eqref{eq:state_measured_ls}, define
\begin{align}
    \hat{X}_0&:=\begin{bmatrix}\hat{x}_0&\hat{x}_1&\cdots&\hat{x}_{N-1}\end{bmatrix},\\
    \hat{X}_+&:=\begin{bmatrix}\hat{x}_1&\hat{x}_2&\cdots&\hat{x}_N\end{bmatrix}.
\end{align}
Collecting the unknowns of the state and output equations into
\begin{align}
    \Theta_x:=\begin{bmatrix}A_0&A_1&B_0&B_1\end{bmatrix},\qquad
    \Theta_y:=\begin{bmatrix}C&D\end{bmatrix},\label{eq:Theta_def}
\end{align}
and defining the regressor matrices
\begin{align}
    \Phi_x:=\begin{bmatrix}
        \hat{X}_0\\
        \hat{X}_0\Delta\\
        U\\
        U\Delta
    \end{bmatrix},\qquad
    \Phi_y:=\begin{bmatrix}\hat{X}_0\\U\end{bmatrix},
    \label{eq:Phi_def}
\end{align}
together with $Y:=\begin{bmatrix}y_0&y_1&\cdots&y_{N-1}\end{bmatrix}$,
the state-equation and output-equation matrices are obtained as the
least-squares solutions
\begin{align}
    \Theta_x^\star
    &= \argmin_{\Theta_x}\|\hat{X}_+ - \Theta_x\Phi_x\|_F^2,
    \label{eq:Theta_x_argmin}\\
    \Theta_y^\star
    &= \argmin_{\Theta_y}\|Y - \Theta_y\Phi_y\|_F^2.
    \label{eq:Theta_y_argmin}
\end{align}
Both problems are decoupled and admit closed-form solutions whenever the
corresponding regressor matrices have full row rank.  In the implementation,
a small ridge term $\lambda I$ with $\lambda>0$ is added to
$\Phi_x\Phi_x^\top$ and $\Phi_y\Phi_y^\top$ in the normal equations for
numerical robustness.
This step is identical to the state-measured least-squares problem
\eqref{eq:state_measured_ls}, except that the true state is replaced by the
smoothed state estimate, so that the regressors are subject to
errors-in-variables effects.  Let $\theta_{\rm LS}^{(r+1)}$ denote the
collection of these least-squares estimates obtained at iteration $r+1$.
To prevent excessively large parameter changes in early iterations,
$\theta_{\rm LS}^{(r+1)}$ is combined with the previous estimate
$\theta^{(r)}$ through the damped update
\begin{align}
    \theta^{(r+1)}=(1-\gamma)\theta^{(r)}+\gamma\theta_{\rm LS}^{(r+1)},
    \qquad 0<\gamma\leq 1,
    \label{eq:damped_update}
\end{align}
where $\gamma$ is the damping factor.  Setting $\gamma=1$ gives an
undamped update, while smaller values provide a more conservative update.

\subsubsection*{Interpretation as alternating minimization}

The overall iterative procedure can be interpreted as deterministic
block-coordinate minimization of a regularized least-squares criterion
defined over the state sequence $(x_k)_{k=0}^{N}$ and the model parameters $\theta$.
For fixed $\theta$, the cost is quadratic in the state trajectory and is
minimized by \eqref{eq:state_smoothing}.  For a fixed state trajectory, the
same cost is quadratic in the LPV matrices and is minimized by the
least-squares refitting step.  The problem is not jointly convex because the
state-equation residual contains bilinear products between the state and the
model matrices.  Therefore, global optimality is not guaranteed.  When the
undamped update $\gamma=1$ is used and each block is solved exactly, the
regularized least-squares cost is non-increasing.  For the damped update
$\gamma<1$ used in the numerical examples, this monotonicity is not
explicitly enforced; the damping is introduced as a numerical stabilization
heuristic.  The procedure is structurally reminiscent of EM-type algorithms
for state-space models \citep{gibson2005}, but is a purely deterministic
alternating least-squares method without probabilistic expectations or
state covariances.

\subsubsection*{Algorithm summary and design parameters}

The complete procedure is summarized in Algorithm~\ref{alg:lpv_id}.
Steps~2 and~3 construct the initial estimate $\theta^{(0)}$ of
Section~\ref{sec:initial_model}, and the subsequent loop alternates the
state-smoothing and parameter-refitting steps described above.
A practical stopping rule is to terminate the iterations when the
improvement of the validation fit and the reduction of the validation RMSE
both remain below prescribed thresholds for a specified number of
consecutive iterations.  In that case, a separate test segment should be
used for final performance reporting.

\begin{algorithm}[t]
\caption{Identification of the sampling-interval-dependent LPV model.}
\label{alg:lpv_id}
\begin{algorithmic}[1]
\Require Input-output data and sampling intervals
$(u_k,y_k,h_k)_{k=0}^{N-1}$, model order $n_x$,
nominal interval $\bar{h}$,
smoothing weight $\alpha$, damping factor $\gamma$,
maximum iteration count $r_{\max}$.
\Ensure Identified matrices
$\theta=(A_0,A_1,B_0,B_1,C,D)$.
\State Compute $\delta_k$ from the sampling instants by
\eqref{eq:hk_def} and \eqref{eq:delta_def}.
\State Identify a standard LTI subspace model, such as an N4SID model,
from the input-output data while ignoring the variation of $h_k$.
\State Embed the LTI model into the LPV parametrization by
\eqref{eq:init_AB0}--\eqref{eq:init_B1} to obtain the initial estimate
$\theta^{(0)}$, with $A_1^{(0)}$ and $B_1^{(0)}$ initialized by the
bilinear approximation of $A_{\rm c}\bar{h}$.
\For{$r=0,1,2,\ldots,r_{\max}-1$}
    \State With $\theta^{(r)}$ fixed, compute
    $\mathcal{A}_k^{(r)}=A_0^{(r)}+A_1^{(r)}\delta_k$ and
    $\mathcal{B}_k^{(r)}=B_0^{(r)}+B_1^{(r)}\delta_k$, and reconstruct the state
    sequence $\hat{X}^{(r+1)}$ by solving \eqref{eq:state_smoothing}.
    \State With $\hat{X}^{(r+1)}$ fixed, compute the least-squares
    estimate $\theta_{\rm LS}^{(r+1)}$ by
    \eqref{eq:Theta_x_argmin}--\eqref{eq:Theta_y_argmin}.
    \State Apply the damped update
    $\theta^{(r+1)}=(1-\gamma)\theta^{(r)}+\gamma\theta_{\rm LS}^{(r+1)}$.
    \If{stopping condition is satisfied}
        \State \textbf{break}.
    \EndIf
\EndFor
\State \textbf{return}~$\theta\leftarrow\theta^{(r+1)}$.
\end{algorithmic}
\end{algorithm}

The main design parameters are the model order $n_x$, the smoothing
weight $\alpha$, and the damping factor $\gamma$.  In practice, $n_x$ can be
chosen from singular-value gaps, cross-validation, or information criteria,
and the role of $\alpha$ is described after \eqref{eq:state_smoothing} and
that of $\gamma$ with \eqref{eq:damped_update}.  In addition, the
scheduling sequence must vary enough to identify $A_1$ and $B_1$ from
$\delta_k x_k$ and $\delta_k u_k$.

\section{Numerical Examples}\label{sec:numerical}

This section evaluates the proposed method in two numerical studies.
Section~\ref{sec:sensitivity} examines, for a noise-free second-order
plant, how the identification accuracy depends on the sampling-interval
variation amplitude.  Section~\ref{sec:noisy} considers a higher-order MIMO
plant under measurement noise and an unmeasured input disturbance,
including the sensitivity to the measurement-noise level.  All iterative
refinements are run for $500$ iterations with the smoothing weight
$\alpha=10^4$, the damping factor $\gamma=0.5$, and the ridge parameter
$\lambda=10^{-8}$.  These values were chosen from preliminary trials and
are fixed across all examples rather than tuned per case; the
validation-based stopping rule of Section~\ref{sec:iterative} is not used.
As in Problem~\ref{prob:id}, the model order is assumed known.  The
identification and validation data are generated independently.
The model quality is evaluated by the fit index
\begin{align}
    {\rm FIT}=100\left(1-\frac{\|Y-\hat{Y}\|_F}{\|Y-\bar{Y}\|_F}\right),
    \label{eq:fit_def}
\end{align}
where $Y$ is the validation output sequence, $\hat{Y}$ is the simulated
output of the identified model on the validation input and sampling
sequence, and $\bar{Y}$ is the matrix of the same dimensions as $Y$ in
which every column equals the channelwise sample mean of $Y$.  In
validation, $\hat{Y}$ is generated by free-run simulation of the
identified model.  The initial validation state is estimated by a
least-squares fit of the free-run output to the first $L$ validation
output samples, and the validation output is not used otherwise.  The FIT
index and RMSE are computed from the free-run outputs over the entire
validation sequence $k=0,\ldots,N_{\rm val}-1$, which thus includes the
first $L$ samples used for initial-state estimation.

In both Sections~\ref{sec:sensitivity} and~\ref{sec:noisy}, the proposed
LPV model is compared with the initial N4SID-LTI model and, as a
diagnostic reference, with a state-measured LPV benchmark.  This
benchmark assumes that the true state sequence
$(x_k)_{k=0}^{N}$ of the identification data is directly available; here
$x_k=x(t_k)$ is expressed in the state coordinates of \eqref{eq:ct_state}
and is generated by the variable-sampling discretization
\eqref{eq:exact_varying_dt} of the continuous-time plant.  The benchmark
identifies the matrices $A_0,A_1,B_0,B_1$ from
$(x_k)_{k=0}^{N}$ and $(u_k,\delta_k)_{k=0}^{N-1}$ as the least-squares solution of
\eqref{eq:state_measured_ls}.  Since the state is expressed in the coordinates of the
plant, the output matrices are set to $C=C_{\rm c}$ and $D=D_{\rm c}$, and
the validation output is simulated from the true initial state of the
validation data rather than from an estimated one.  Although not
implementable from input-output data alone, this benchmark serves as a
practical upper-bound reference.  It indicates the best validation performance
achievable by the affine LPV structure when the state trajectory, the
output matrices, and the initial state are all known exactly.
Accordingly, the gap between the proposed model and the benchmark measures
the combined cost of reconstructing the state from input-output data and of
estimating $C$, $D$, and the initial validation state, whereas the gap
between the benchmark and a perfect fit reflects the affine truncation
error $e_k$ in \eqref{eq:lpv_state} and, where present, the disturbances
$\eta_k$ and $v_k$.

\subsection{Sensitivity to the sampling-interval variation amplitude}\label{sec:sensitivity}

Consider the continuous-time plant
\begin{align}
    A_{\rm c}=\begin{bmatrix}0&1\\-4&-1\end{bmatrix},\quad
    B_{\rm c}=\begin{bmatrix}0\\4\end{bmatrix},\quad
    C_{\rm c}=\begin{bmatrix}1&0\end{bmatrix},\quad
    D_{\rm c}=0,
    \label{eq:siso_plant}
\end{align}
which has two states, one input, and one output, and is controllable and
observable.  The sampling intervals are generated as
\begin{align}
    h_k=0.05(1+a\rho_k),\qquad \rho_k\in[-1,1],
    \label{eq:hk_random}
\end{align}
where $\rho_k$ are independent samples from the uniform distribution on
$[-1,1]$.  The corresponding observed range is
$h_{\min}=0.05(1-a)$ and $h_{\max}=0.05(1+a)$, and we adopt the sample
median of the identification intervals as the nominal sampling
interval $\bar{h}$ introduced in Section~\ref{sec:formulation}.  It is
approximately equal to the midpoint $0.05$, so that the normalized
variation $\delta_k$ is approximately confined to $[-a,a]$ and centered
around zero.
The input $u_k$ is independently drawn from the
standard normal distribution and held constant over $[t_k,t_{k+1})$.  The
identification and validation data are generated with the same value of
$a$ from independent realizations of $\rho_k$ and $u_k$, with lengths
$N_{\rm id}=3000$ and $N_{\rm val}=1500$, respectively.  The normalized
variation $\delta_k$ of the validation intervals is computed with the
$\bar{h}$ determined from the identification data.  The initial-state
estimation window is $L=100$.
No measurement noise or input disturbance is added in this subsection, so
that the result reflects the effect of sampling-interval variation.  In
connection with Assumption~\ref{ass:plant_sampling}, this
example has $\delta_{\max}\approx a$ and $\|A_{\rm c}\|_2\approx 4.13$, so
$\|A_{\rm c}\|_2\bar{h}\delta_{\max}$ ranges from about $0.010$ at $a=0.05$ to
about $0.19$ at $a=0.90$.  Even the largest tested variation thus remains in
the regime where the first-order expansion of
Section~\ref{sec:formulation} is meaningful.  The LPV columns of
Table~\ref{tab:siso_sensitivity} refer to the model after the iterative
refinement of Algorithm~\ref{alg:lpv_id}.

\begin{table}[t]
\centering
\caption{SISO sampling-variation study: validation FIT [\%] and RMSE
of the N4SID-LTI and proposed LPV models.}
\label{tab:siso_sensitivity}
\begin{tabular}{cccc}
\toprule
$a$ & LTI FIT & LPV FIT & LPV RMSE \\
\midrule
0.05 & 96.996 & 99.996 & $1.24\times10^{-5}$ \\
0.10 & 93.717 & 99.985 & $4.77\times10^{-5}$ \\
0.20 & 87.616 & 99.949 & $1.68\times10^{-4}$ \\
0.30 & 80.166 & 99.848 & $5.09\times10^{-4}$ \\
0.50 & 72.410 & 99.573 & $1.43\times10^{-3}$ \\
0.70 & 51.319 & 99.184 & $2.75\times10^{-3}$ \\
0.90 & 42.930 & 98.612 & $4.68\times10^{-3}$ \\
\bottomrule
\end{tabular}
\end{table}

\begin{figure}[t]
\centering
\includegraphics[width=0.45\textwidth]{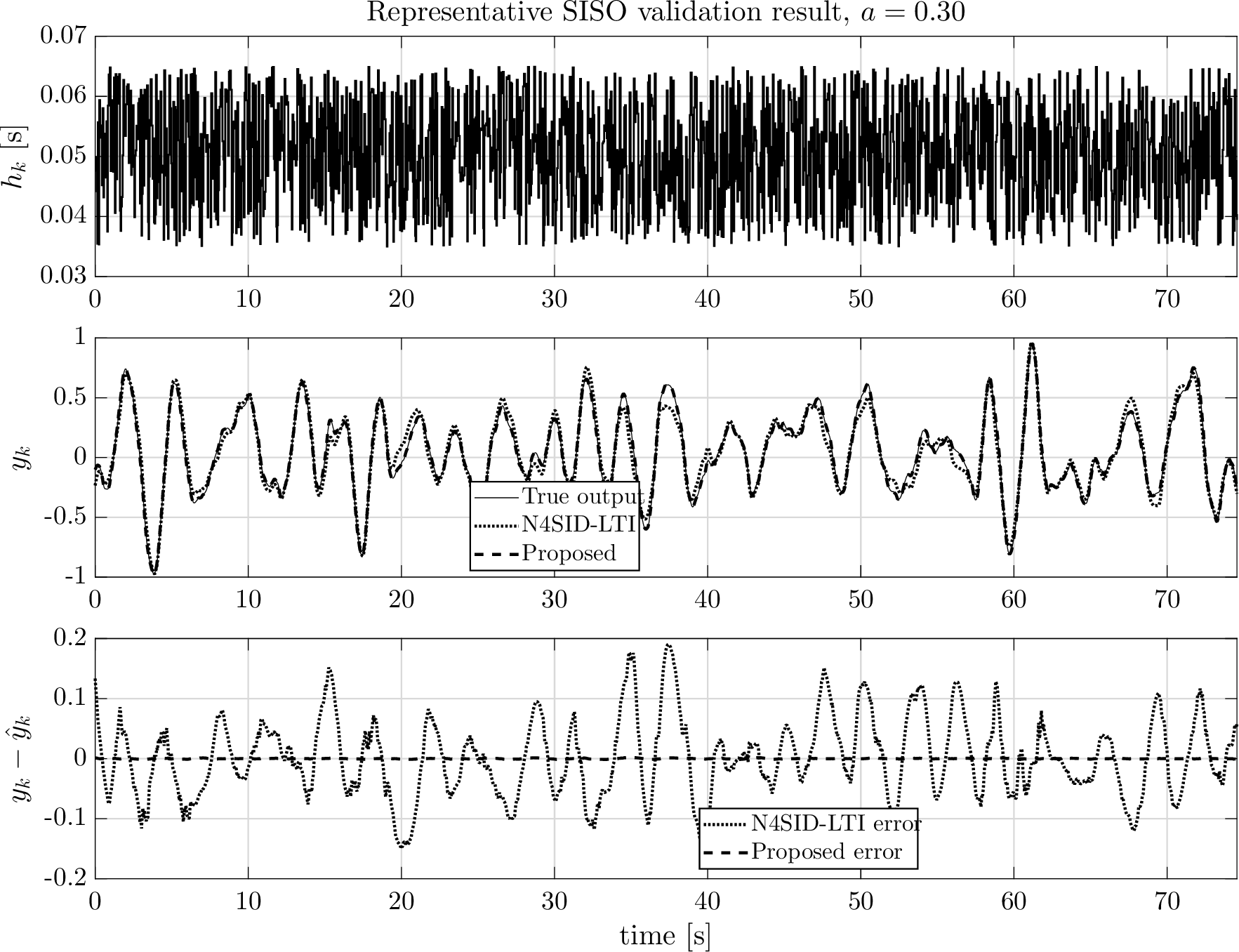}
\caption{
Representative validation result for the SISO sampling-variation study
at $a=0.30$.  Top: aperiodic sampling-interval sequence.  Middle: true
output and simulated outputs of the N4SID-LTI and proposed LPV models.
Bottom: corresponding output errors.}
\label{fig:siso_waveform}
\end{figure}

Table~\ref{tab:siso_sensitivity} shows that the validation FIT of the
N4SID-LTI model decreases rapidly as $a$ increases, from $96.996\%$ at
$a=0.05$ to $42.930\%$ at the large-variation stress case $a=0.90$,
whereas the proposed LPV model maintains a FIT above $98.6\%$ over the
entire range.  Figure~\ref{fig:siso_waveform} visualizes the representative
case $a=0.30$: the fixed LTI response exhibits a visible output mismatch
that is largely removed by the proposed LPV model.  For this case, the
proposed LPV model achieves a FIT of $99.848\%$ with RMSE
$5.09\times10^{-4}$, nearly identical to the $99.849\%$ and
$5.04\times10^{-4}$ of the state-measured LPV benchmark, which serves as
the upper-bound reference for the affine LPV structure and remains below
$100\%$ because of the affine truncation error $e_k$ in
\eqref{eq:lpv_state}.  These results show that the proposed method
recovers the variable-sampling behavior over the entire tested range.

\subsection{Identification under input disturbance and measurement noise}\label{sec:noisy}

Consider the three-state, two-input two-output continuous-time plant
\begin{align}
    A_{\rm c}&=
    \begin{bmatrix}
        0&1&0\\
        0&0&1\\
        -8&-14&-7
    \end{bmatrix},&
    B_{\rm c}&=
    \begin{bmatrix}
        0&0\\
        1&0\\
        0&1
    \end{bmatrix},\notag\\
    C_{\rm c}&=
    \begin{bmatrix}
        1&0&0\\
        0&1&0
    \end{bmatrix},&
    D_{\rm c}&=0_{2\times 2},
    \label{eq:numerical_plant}
\end{align}
which is controllable and observable.  The sampling intervals of both the
identification and validation data are generated by \eqref{eq:hk_random}
with $a=0.35$, and the nominal sampling
interval $\bar{h}$ is the sample median of the identification intervals,
approximately equal to the midpoint $0.05$ of the range
$[0.05(1-a),0.05(1+a)]$.  For this plant,
$\|A_{\rm c}\|_2\approx 17.6$, so
$\|A_{\rm c}\|_2\bar{h}\delta_{\max}\approx 0.31$ at $\delta_{\max}\approx a=0.35$,
which stresses the affine approximation of
Assumption~\ref{ass:plant_sampling} more than in the SISO study of
Section~\ref{sec:sensitivity}.  The identification and validation data
lengths are $N_{\rm id}=3000$ and $N_{\rm val}=1200$, respectively.  Each
component of the input $u_k\in\Real^2$ is independently drawn from the
standard normal distribution and held constant over $[t_k,t_{k+1})$.  The
data-generating system is
\begin{align}
    x_{k+1}&=A_{\rm d}(h_k)x_k+B_{\rm d}(h_k)(u_k+\eta_k),
    \label{eq:numerical_input_disturbance}\\
    y_k&=Cx_k+D(u_k+\eta_k)+v_k,
    \label{eq:numerical_noisy_output}
\end{align}
where $\eta_k\in\Real^2$ is an unmeasured input disturbance and $v_k$ is
measurement noise.  Each component of $\eta_k$ and $v_k$ is independently
drawn from a zero-mean Gaussian distribution with standard deviations
$\sigma_\eta=10^{-2}$ and $\sigma_v=10^{-2}$, respectively.
Since $D=D_{\rm c}=0$ in this example, the term $D\eta_k$ in
\eqref{eq:lpv_output} vanishes and the input disturbance $\eta_k$ acts
only through the state equation.  The identification algorithm is given
only $u_k$, $y_k$, and $h_k$.

In this example, the FIT index \eqref{eq:fit_def} and the RMSE are
computed with respect to two reference outputs $Y$: the nominal-clean
output, which is the plant output with neither input disturbance nor
measurement noise, and the noisy output, which contains both.  The
nominal-clean fit is the main indicator of deterministic plant-model
accuracy, while the noisy-output fit is limited by the measurement-noise
floor.  In both cases the initial validation state is estimated from the
first $L=120$ samples of the noisy output.  The validation output is not
used for state smoothing or parameter refitting.

The proposed LPV model is compared with the initial N4SID-LTI model, the
state-measured LPV benchmark defined at the beginning of
Section~\ref{sec:numerical}, the initial LPV model (the
Tustin-initialized estimate $\theta^{(0)}$ obtained at Step~3 of
Algorithm~\ref{alg:lpv_id}), and an LPV-ARX baseline.  The proposed LPV
model is the final estimate after the iterative refinement.  In the MATLAB
implementation of the N4SID initialization, the \texttt{iddata} sample
time is set to $\bar{h}$ and the direct-feedthrough term is allowed.

As an input-output baseline, we use the $\delta_k$-affine LPV-ARX model
\begin{align}
    y_k&=\sum_{i=1}^{n_a}\bigl(A_{i,0}+A_{i,1}\delta_{k-1}\bigr)y_{k-i}
    \notag\\
    &\quad+\sum_{j=0}^{n_b}\bigl(B_{j,0}+B_{j,1}\delta_{k-1}\bigr)u_{k-j},
    \label{eq:lpv_arx}
\end{align}
with $n_a=3$ and $n_b=2$, in which every coefficient matrix is
an affine function of $\delta_{k-1}$, the normalized variation of the
interval that governs the transition from $x_{k-1}$ to $x_k$.  The
coefficient matrices are identified by one-shot linear least squares on
the equation error, that is, the one-step-ahead prediction error
\citep{toth2010}.  In validation the model is simulated in free run with
the first $n_a$ measured validation outputs as initial conditions, and
the FIT index and RMSE are computed as for the other models.

\begin{table}[t]
\centering
\caption{MIMO identification under measurement noise ($\sigma_v=10^{-2}$)
and input disturbance ($\sigma_\eta=10^{-2}$): FIT [\%] and nominal-clean
RMSE in units of $10^{-3}$.  LPV-ARX: model \eqref{eq:lpv_arx}.
State-meas.\ LPV: state-measured LPV benchmark of
Section~\ref{sec:numerical}.}
\label{tab:mimo_noisy}
\begin{tabular}{@{}lccc@{}}
\toprule
Model & Nom.\ FIT & Noisy FIT & RMSE \\
\midrule
N4SID-LTI          & 79.35 & 77.99 & 26.3  \\
LPV-ARX            & 87.44 & 85.08 & 16.0  \\
Initial LPV        & 97.14 & 91.53 & 3.65  \\
Proposed LPV       & 99.18 & 92.02 & 1.04  \\
State-meas.\ LPV   & 99.83 & 92.07 & 0.219 \\
\bottomrule
\end{tabular}
\end{table}

Table~\ref{tab:mimo_noisy} shows the effect of each identification stage.
The fixed N4SID-LTI model ($A_1=B_1=0$) achieves a nominal-clean FIT of
only $79.35\%$.  The LPV-ARX model, which shares the scheduling variable
and the affine dependence but is identified by a single linear regression
without state reconstruction, achieves $87.44\%$.  This confirms the value
of the affine scheduling structure, but the input-output regression does
not exploit the state-space structure and is limited by the error
accumulation inherent in free-run ARX simulation.  The initial LPV model
already achieves $97.14\%$.  Since it shares $A_0$, $B_0$, $C$, and $D$
with the N4SID-LTI model and differs only by $A_1^{(0)}$ and
$B_1^{(0)}$ of \eqref{eq:init_A1}--\eqref{eq:init_B1}, this gain shows
that the Tustin approximation of $A_{\rm c}\bar{h}$ captures most of the
sampling-interval dependence from the LTI realization alone.  The proposed LPV model further improves the
nominal-clean FIT to $99.18\%$, and its noisy-output FIT ($92.02\%$) is
close to that of the state-measured LPV benchmark ($92.07\%$).  The larger
part of the improvement from LPV-ARX to the proposed LPV model is
thus already obtained by the state-space LPV structure with the
sampled-data-based initialization, while the state-sequence
reconstruction and iterative refitting reduce the nominal RMSE by a
further factor of about $3.5$.  The residual gap to the benchmark
reflects state-reconstruction error, the estimation of $C$, $D$, and the
initial state, and noise-induced bias.

To examine the sensitivity to measurement noise, $\sigma_v$ is varied over
$\{0,10^{-2},3\times10^{-2}\}$ while fixing the
input-disturbance level at $\sigma_\eta=10^{-2}$.  The proposed LPV model
maintains high nominal-clean accuracy over this range: its FIT is
$99.67\%$, $99.18\%$, and $98.13\%$, respectively (the middle value being
the representative case of Table~\ref{tab:mimo_noisy}).  The corresponding
LPV-ARX values are $88.75\%$, $87.44\%$, and $78.67\%$, so the proposed
method outperforms the one-shot input-output regression by
approximately $11$--$19$ percentage points.

As a reference comparison, a continuous-time indirect method is also
evaluated.  An initial continuous-time state-space model is estimated by
a simulation-focus prediction-error method (\texttt{ssest} in MATLAB,
applied to the data with the nominal sample time $\bar{h}$).  The
estimates $(A_{\rm c},B_{\rm c},C_{\rm c},D_{\rm c})$ are then refined by
nonlinear least squares (\texttt{lsqnonlin}) that minimizes the free-run
output residual under exact variable-step ZOH discretization, and the
refined estimates are converted to affine LPV coefficients via
\eqref{eq:Ad_taylor}--\eqref{eq:Bd_taylor}.  The FIT values reported
below for this method are those of the converted affine LPV model.
For $\sigma_v\le 10^{-2}$, the proposed method outperforms the CT indirect
method ($99.67\%$ vs.\ $98.97\%$ at $\sigma_v=0$, and $99.18\%$ vs.\
$98.69\%$ at $\sigma_v=10^{-2}$), and at $\sigma_v=3\times10^{-2}$ the two
are comparable ($98.13\%$ vs.\ $98.28\%$).  Since the true plant belongs
to the model class of the CT indirect method, the advantage is plausibly
due to fitting the affine coefficients directly by least squares rather
than obtaining them through the Taylor conversion
\eqref{eq:Ad_taylor}--\eqref{eq:Bd_taylor}, and possibly to local minima
of the nonlinear least-squares refinement.  At a higher noise level,
$\sigma_v=5\times10^{-2}$, the CT indirect method ($97.77\%$) is more
accurate than the proposed LPV model ($90.41\%$); we conjecture that this
is because the fixed smoothing weight $\alpha$ does not adapt to the
noise level.  Within $\sigma_v\le 3\times10^{-2}$, the directly identified
affine LPV model thus attains accuracy comparable to or higher than that
of the indirect route, without continuous-time estimation or model
conversion.

\section{Conclusion}\label{sec:conclusion}

This paper proposed a method that identifies, from aperiodically sampled
input-output data, a control-oriented model: a discrete-time affine LPV
state-space model scheduled by the normalized sampling-interval variation.
Its affine structure, inherited from the first-order expansion of the
zero-order-hold discretization, makes the identified model directly usable
in LMI-based analysis and synthesis without conversion from a
continuous-time model.
The method derives an initial LPV model, including approximate first-order
terms, from an LTI subspace model through sampled-data relations, and then
alternates least-squares state reconstruction and parameter refitting.

In the numerical examples, the proposed LPV model kept the validation FIT
above $98.6\%$ for sampling-interval variations of up to $\pm 90\%$ of the
nominal interval, where a fixed LTI model dropped to about $43\%$.  Under
measurement noise and an unmeasured input disturbance, it attained a FIT
of $99.18\%$ with respect to the undisturbed, noise-free output, against
$87.44\%$ for an LPV-ARX baseline,
and approached the state-measured LPV benchmark ($99.83\%$), a practical
upper-bound reference.  Most of this gain was already obtained by the
initial LPV model, and the iterative refinement further reduced the
corresponding RMSE by a factor of about $3.5$.  The accuracy was also
comparable to or higher than that of a continuous-time indirect method
requiring such a conversion, except at the strongest noise level tested.

Future work includes higher-order scheduling bases,
which can be incorporated by appending regressor blocks to the refitting
step.  Another direction is the extension to closed-loop data and
open-loop unstable plants.  A natural next step is control design with the
identified model, including the case of stochastic sampling, while
accounting for the truncation and estimation errors.

\section*{Funding}
This work was supported by JSPS KAKENHI Grant Number JP26K07555.

\section*{Declaration of generative AI and AI-assisted technologies in the writing process}
During the preparation of this work, the authors used Claude (Anthropic) in
order to improve the readability and language of the manuscript and to check
the dimensional and notational consistency of the mathematical expressions.
After using this tool, the authors reviewed and edited the content as needed
and take full responsibility for the content of the published article.  All
technical content, research methodology, results, and conclusions are the
authors' own work.

\end{document}